%% file: main.tex
\documentclass[conference,compsoc]{IEEEtran}
\IEEEoverridecommandlockouts

\usepackage{amsmath,amssymb,amsfonts}
\usepackage{algorithmic}
\usepackage{graphicx}
\usepackage{textcomp}
\usepackage{xcolor}
\usepackage{hyperref}
\usepackage{biblatex}
\usepackage{numprint}

\usepackage{booktabs}
\usepackage{tabularx}
\usepackage[nolist]{acronym}
\usepackage{xspace}
\usepackage{csquotes}

\newcommand{\HAL}{\texttt{HAL}\xspace}

\usepackage{tabularx}
\newcolumntype{Y}{>{\centering\arraybackslash}X}
\newcolumntype{L}[1]{>{\raggedright\let\newline\\\arraybackslash\hspace{0pt}}m{#1}}
\newcolumntype{C}[1]{>{\centering\let\newline\\\arraybackslash\hspace{0pt}}m{#1}}
\newcolumntype{R}[1]{>{\raggedleft\let\newline\\\arraybackslash\hspace{0pt}}m{#1}}

\def\BibTeX{{\rm B\kern-.05em{\sc i\kern-.025em b}\kern-.08em
    T\kern-.1667em\lower.7ex\hbox{E}\kern-.125emX}}

\def\orcid#1{\kern.08em\href{https://orcid.org/#1}{\protect\includegraphics[keepaspectratio,width=0.7em]{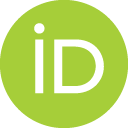}}}

\newcommand{\etal}{et~al.\@\xspace}

\usepackage{circledtext}
\circledtextset{boxfill=red}
\circledtextset{boxtype=o}
\circledtextset{boxcolor=red}
\circledtextset{charf=\bfseries\Large}
\circledtextset{charcolor=white}
\circledtextset{charshrink=0.6}
\circledtextset{resize=real}
\circledtextset{width=0.8em}

\input{acronyms}

\begin{document}

\title{HAL -- An Open-Source Framework for Gate-Level Netlist Analysis
\thanks{Funded by the Deutsche Forschungsgemeinschaft (DFG, German Research Foundation) under Germany´s Excellence Strategy - EXC 2092 CASA – 390781972.}
}

\author{\IEEEauthorblockN{
Julian Speith\IEEEauthorrefmark{1}\orcid{0000-0002-8408-8518},
Jörn Langheinrich\IEEEauthorrefmark{1}\orcid{0000-0002-8583-5503},
Marc Fyrbiak\IEEEauthorrefmark{1}\orcid{0000-0002-4266-7108},
Max Hoffmann\IEEEauthorrefmark{1}\orcid{0000-0001-9072-611X},
Sebastian Wallat\IEEEauthorrefmark{1}\orcid{0000-0002-7429-1002},\\
Simon Klix\IEEEauthorrefmark{1}\orcid{0000-0002-9369-2901},
Nils Albartus\IEEEauthorrefmark{1}\orcid{0000-0003-2449-1134},
René Walendy\IEEEauthorrefmark{1}\orcid{0000-0002-5378-3833},
Steffen Becker\IEEEauthorrefmark{1}\IEEEauthorrefmark{2}\orcid{0000-0001-7526-5597},
Christof Paar\IEEEauthorrefmark{1}\orcid{0000-0001-8681-2277}}
\IEEEauthorblockA{
\IEEEauthorrefmark{1}Max Planck Institute for Security and Privacy (MPI-SP)\\
\IEEEauthorrefmark{2}Ruhr University Bochum (RUB)}
}

\maketitle

\input{sections/00_abstract}


\input{sections/01_intro}

\input{sections/02_artifact}
\input{sections/03_impact}

\input{sections/04_conclusion}

\printbibliography

\end{document}

%% file: acronyms.tex
\begin{acronym}
    \acro{36C3}{36th Chaos Communication Congress}

    \acro{API}{application programming interface}
    \acro{ASIC}{application-specific integrated circuit}

    \acro{CI}{continuous integration}

    \acro{FHE}{fully homomorphic encryption}
    \acro{FPGA}{field-programmable gate array}
    \acro{FSIC}{Free Silicon Conference}
    \acro{FSM}{finite state machine}

    \acro{GNN}{graph neural network}
    \acro{GUI}{graphical user interface}

    \acro{HRE}{hardware reverse engineering}
    \acro{HW}{hardware}

    \acro{IC}{integrated circuit}
    \acro{IP}{intellectual property}

    \acro{MPI-SP}{Max Planck Institute for Security and Privacy}
    
    \acro{RE}{reverse engineering}
    \acro{RUB}{Ruhr University Bochum}

    \acro{SEM}{scanning electron microscope}
    \acro{SMT}{satisfiability modulo theories}
    \acro{SoC}{system on a chip}

    \acro{WSL}{Windows Subsystem for Linux}
\end{acronym}

%% file: sections/00_abstract.tex
\begin{abstract}
HAL is an open-source framework for gate-level netlist analysis, an integral step in hardware reverse engineering. It provides analysts with an interactive GUI, an extensible plugin system, and APIs in both C++ and Python for rapid prototyping and automation. In addition, HAL ships with plugins for word-level modularization, cryptographic analysis, simulation, and graph-based exploration. Since its release in 2019, HAL has become widely adopted in academia, industry, government, and teaching. It underpins at least 23 academic publications, is taught in hands-on trainings, conference tutorials, and university classes, and has collected over 680 stars and 86 forks on GitHub. By enabling accessible and reproducible hardware reverse engineering research, HAL has significantly advanced the field and the understanding of real-world capabilities and threats.
\end{abstract}

%% file: sections/01_intro.tex
\section{Introduction}

Our \HAL framework is an essential tool in a typical \ac{HRE} workflow.
\Ac{HRE} is vital for hardware assurance, especially when components manufactured offshore are deployed in high-stakes applications.
Today, hardware manufacturing is frequently outsourced to untrusted facilities across the globe, involving numerous stakeholders with divergent economic incentives and competing government interests.
Consequently, incorporating such technologies into critical infrastructure and defense applications necessitates rigorous validation.
\ac{HRE} plays a vital role in enabling such hardware validation~\cite{DBLP:journals/jetc/BoteroWLRMGATWF21}.
Other benign applications of \ac{HRE} include detecting \ac{IP} infringement, competitor analysis, and the search for hardware vulnerabilities.
However, malicious entities may also use \ac{HRE} to conduct \ac{IP} theft, find and exploit zero-day vulnerabilities, or reverse engineer and break proprietary cryptographic protocols.
Therefore, it is important to study \ac{HRE} not only for hardware assurance, but also to better understand real-world attacker capabilities.

\Ac{HRE} generally comprises two main phases:
First, a gate-level netlist is recovered from an \ac{IC}.
For \acp{ASIC}, this often requires a destructive process involving expensive machinery and skilled personnel~\cite {DBLP:conf/ivsw/FyrbiakSKWERP17,DBLP:conf/dac/TorranceJ11}.
In contrast, for \acp{FPGA}, netlist recovery mainly involves decoding the proprietary format of its configuration bitstream file~\cite{DBLP:conf/iscas/NarwariyaPDA25}.
In the second phase, netlist analysis, the reverse engineer faces a (potentially incomplete) netlist containing hundreds of thousands of gates that lack hierarchy or structural information~\cite{DBLP:journals/jce/AzrielSAGMP21,DBLP:conf/ccs/KlixASSVWLLKSHP24}.
Especially for security-critical components, designers may have employed obfuscation techniques to impede reverse engineering.

To date, researchers and practitioners in the \ac{HRE} field have primarily depended on their own collections of scripts, proprietary internal software, or adaptations of hardware design tools for reverse engineering.  
This fragmentation significantly hinders reproducibility and the sharing of results within the community.
Moreover, due to the sensitive nature of \ac{HRE}, many such efforts take place behind closed doors.
Consequently, the true real-world capabilities of entities conducting such analyses remain largely unknown.

This is where our open-source \HAL framework~\cite{hal_github} comes into play.
To address the aforementioned issues, we positioned \HAL to
\begin{itemize}
    \item \textbf{[accessibility]} significantly lower the entry barrier for new researchers entering the domain of \ac{HRE},
    \item \textbf{[reproducibility]} enhance reproducibility in \ac{HRE} research and support artifact evaluation efforts, 
    \item \textbf{[attacker capabilities]} and facilitate a better understanding and assessment of real-world attacker capabilities.
\end{itemize}

\HAL addresses a crucial step in \ac{HRE}, namely netlist analysis.
It enables in-depth analysis of arbitrary gate-level netlists, whether recovered from an \ac{ASIC} or an \ac{FPGA}.
\HAL not only ships with a suite of algorithms specifically designed for critical reverse engineering tasks but also provides means for rapid prototype development and hypothesis testing, supporting analysts in their mission to uncover a netlist's high-level functionality.
To this end, \HAL abstracts away recurring tasks such as netlist parsing and offers interfaces for straightforward netlist exploration---whether through a \ac{GUI}, Python scripts, or specialized plugins.
Performance is robust thanks to its highly optimized C++ core, while flexibility is ensured through a built-in Python \ac{API}, modularity via a dynamic plugin system, and stability is maintained by a rigorous test suite.
We envision \HAL to be the netlist equivalent of software reverse engineering tools like IDA Pro and Ghidra.

\subsubsection*{Impact}
\HAL has been publicly available on GitHub~\cite{hal_github} since 2019.
Over the years, it has amassed more than 680 stars and has been forked 86 times.
Today, \HAL is used to teach university classes attended by over 200 students and to deliver workshops and tutorials to approximately 150 academics and practitioners.
A total of 23 academic publications have built upon \HAL, and the two foundational publications~\cite{DBLP:journals/tdsc/FyrbiakWSHHWWTP19,DBLP:conf/cf/WallatA00EFDMP19} have been cited more than 150 times.
To the best of our knowledge, \HAL is the only publicly maintained tool for netlist reverse engineering; we are not aware of any commercial equivalents.
However, these metrics cannot fully capture \HAL's impact on the community, as many stakeholders operate behind closed doors, and even companies often avoid being publicly associated with \ac{HRE}.

\subsubsection*{Case Study}
\begin{figure}[htbp]
    \includegraphics[width=\linewidth]{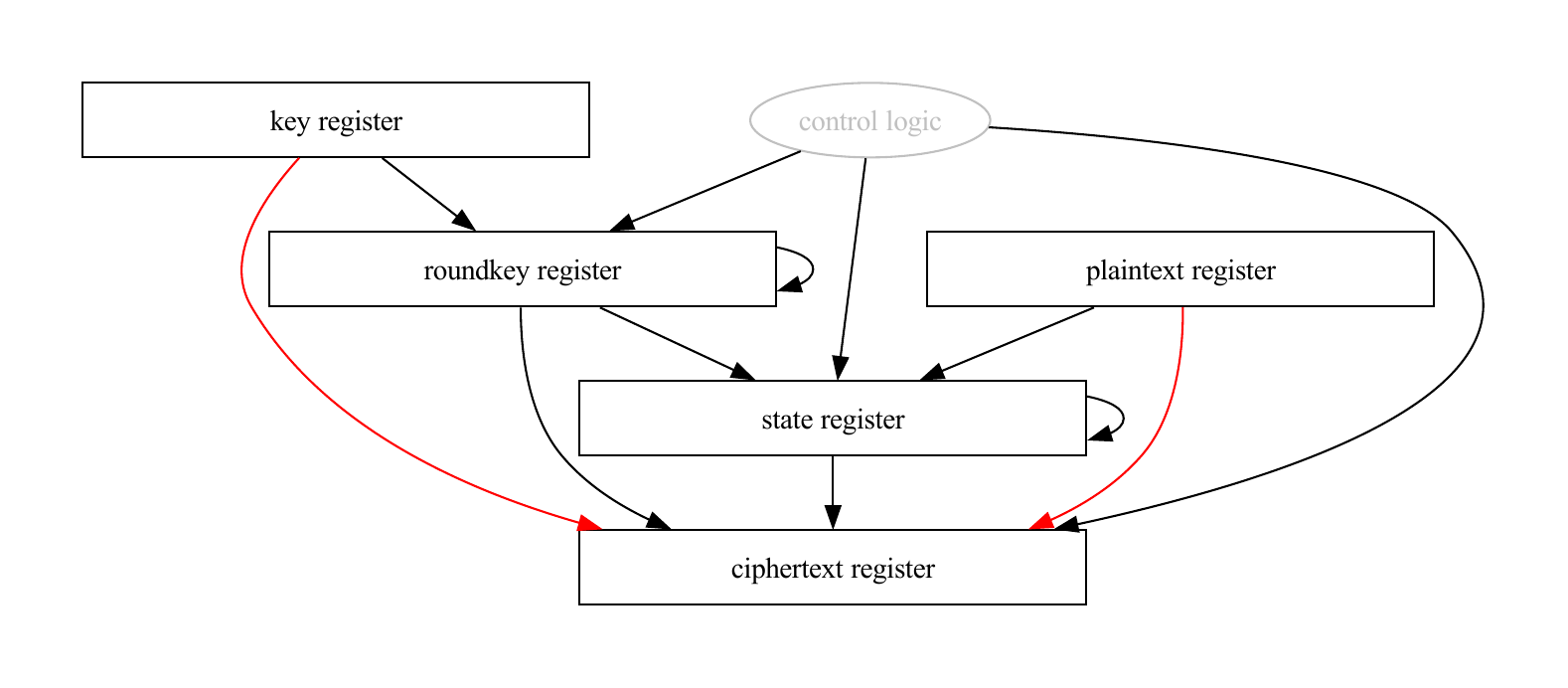}
    \caption{Automatically recovered registers of the cryptographic implementation and their interconnections.
    The suspicious connections are highlighted in red.
    Control logic was omitted here for readability.}
    \label{hal::figure::dataflow}
\end{figure}

To better understand \HAL's capabilities and its importance to hardware assurance, let us look at a Trojan detection case study.\footnote{The corresponding \HAL project can be found in the \texttt{examples} folder that is part of the \HAL GitHub repository~\cite{hal_github}. We encourage the reader to import that project into \HAL and then follow the instructions in the \HAL Wiki at \url{https://github.com/emsec/hal/wiki/Crypto-Trojan}.}
We assume that a Trojan was inserted into a cryptographic accelerator that is part of a larger \ac{SoC}.
Using one of \HAL's automated algorithms, we can easily locate the cryptographic sub-circuit within the netlist.
Using another technique that recovers word-level registers and their interconnections from the unstructured netlist (also known as the \enquote{sea of gates}), we can generate the dataflow graph depicted in \autoref{hal::figure::dataflow}.

Within this graph, we can now look for suspicious connections between registers.
Here, neither the key register nor the plaintext registers should be connected to the ciphertext output.
Such connections already hint at a Trojan that leaks the secret key when a certain plaintext is applied.
However, verifying this assumption requires manual analysis of the suspicious connections, which is straightforward using the netlist graph view within the \HAL \ac{GUI} shown in \autoref{hal::figure::gui}.

%% file: sections/02_artifact.tex
\section{Artifact Description}
\HAL is a technology-agnostic netlist analysis framework that can be used to reverse engineer the high-level functionality of (parts of) a gate-level netlist from any \ac{ASIC} or \ac{FPGA}.
\HAL supports installation on Ubuntu 22.04 and 24.04, \mbox{macOS}, and Windows through the \ac{WSL}. 
The framework consists of three main components: the \HAL core, a \ac{GUI}, and a plugin system.

\subsection{Core}
The \HAL core handles loading netlists and gate libraries, managing projects, and provides a C++ and a Python \ac{API} to interact with the netlist.
To load a Verilog or VHDL netlist into \HAL, a gate library (also known as \textit{standard cell library}) is required.
This library specifies the connectivity and functionality of the combinational and sequential gates in the netlist.
Typically, the gate library is generated during the hardware reverse engineering process and is provided in the widely-used \textit{liberty} file format common in hardware design.
Upon importing a netlist into \HAL, the core creates a project and stores the netlist in a custom file format optimized for efficiency.
Projects can be exported and imported, enabling seamless sharing of reverse engineering progress between workstations.
The core provides individual classes representing the netlist and its gates, nets, and modules, facilitating straightforward access to all netlist elements.
Furthermore, it features Boolean function integration, supporting evaluation, manipulation, and simplification of functions as well as \ac{SMT} solving.

\begin{figure*}
    \centering
    \includegraphics[width=\textwidth]{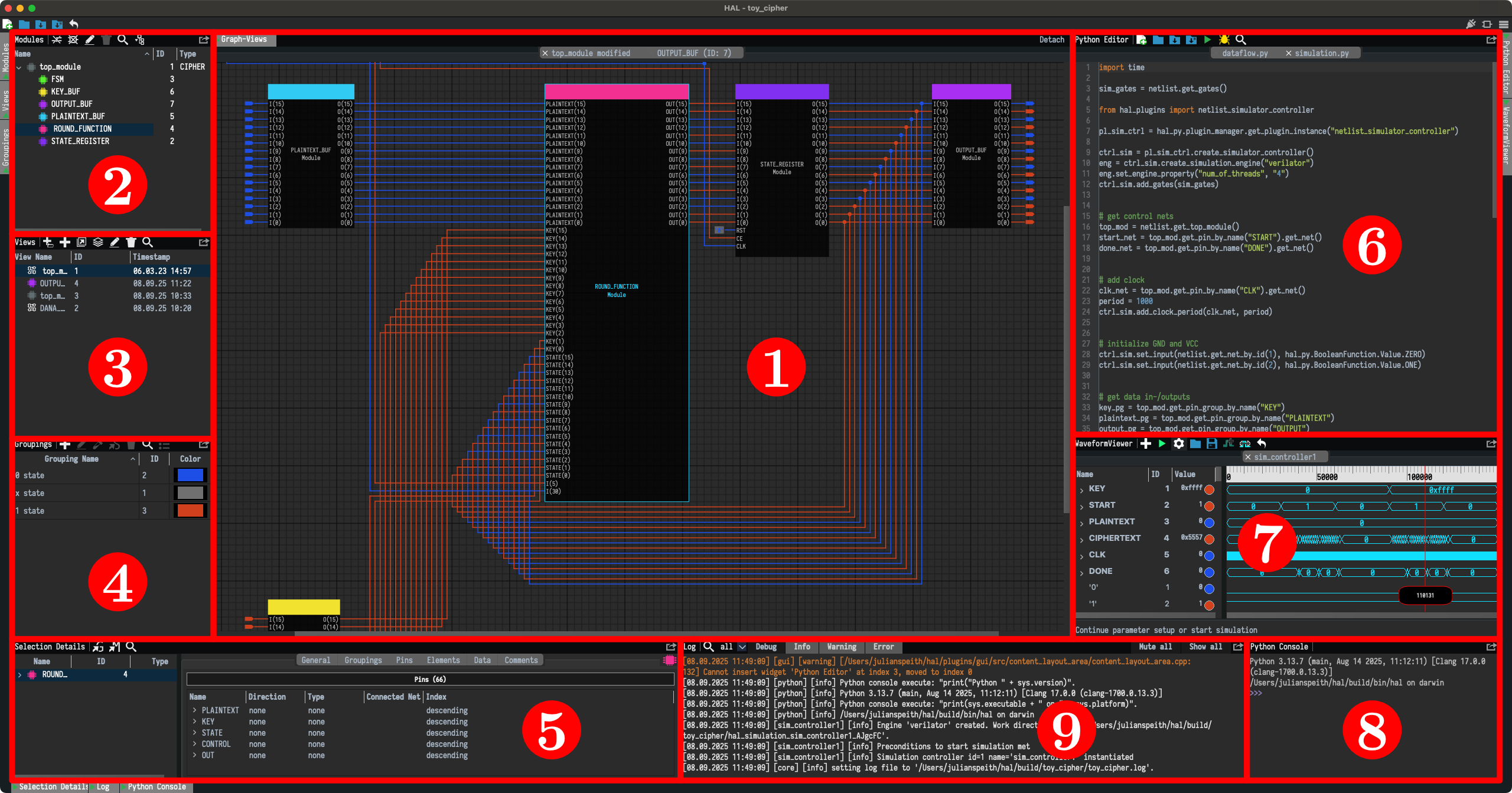}
    \caption{The \HAL \acs{GUI}.}
    \label{hal::figure::gui}
\end{figure*}

\subsection{GUI}
The \HAL \ac{GUI} (see \autoref{hal::figure::gui}) centers around the graph view~\circledtext{1} visualizing the netlist under analysis.
Multiple views allow for different netlist excerpts to be displayed concurrently for efficient manual exploration.
Users can reintroduce hierarchy by grouping gates into modules.
The cone view feature enables progressive exploration starting from a selected object (gate or module), unfolding preceding and succeeding objects using the arrow keys.

Surrounding the graph view, \HAL provides a collection of widgets to display (and operate on) different aspects of the netlist.
In the modules widget~\circledtext{2}, the analyst can explore the module hierarchy of the netlist and toggle a list of all gates and nets contained within the respective modules.
The views widget~\circledtext{3} enables managing existing graph views of the netlist.
In the grouping widget~\circledtext{4}, colored groupings of gates, nets, and modules can be created to support visual analysis.
The selection details widget~\circledtext{5} lists all currently selected gates, nets, and modules.
Furthermore, it provides details on the selected objects such as their name, ID, functions, input and output pin mapping, and more.
The \ac{GUI} also features a Python editor~\circledtext{6} and console~\circledtext{8} for semi-automated netlist exploration and prototyping of automated reverse engineering algorithms.
The waveform viewer~\circledtext{7} enables dynamic analysis of the netlist over time.
The simulation state can be represented in the graph view~\circledtext{1} by coloring nets according to their value at a user-specified point in time.
A logging widget~\circledtext{9} displays important status messages as well as warnings and errors.

\subsection{Plugin System}
\label{hal::subsec::plugins}
The powerful C++ \ac{API} makes extending \HAL straightforward.
Plugins can augment both the core and the \ac{GUI}.

\HAL ships with several off-the-shelf plugins:
The \textit{HAWKEYE} plugin automatically scans the given netlist for implementations of symmetric cryptographic algorithms.
When HAWKEYE finds a candidate for such an implementation, it attempts to automatically identify the implemented cipher by analyzing its S-boxes.
HAWKEYE was initially presented at IACR Crypto 2024~\cite{DBLP:conf/crypto/LeanderPSS24}.

The \textit{dataflow analysis} plugin enables the automated recovery of high-level registers from the unstructured sea of gates. 
It was first presented as DANA~\cite{DBLP:journals/tches/AlbartusHTAP20} at IACR CHES 2020.
DANA significantly contributes to reconstructing word-level modules within a netlist, which is an essential step toward achieving a high-level understanding of a netlist.

In addition to reconstructing registers, the \textit{module identification} plugin identifies word-level arithmetic operations such as additions, subtractions, constant multiplications, and counters within a netlist.
\ac{SMT} solving is used to verify the functionality of the analyzed sub-circuit.

The bit order of recovered word-level structures is usually unknown, since such information is lost during hardware synthesis.
Thus, the \textit{bit-order propagation} plugin uses anchor points such as arithmetic operations and shift registers that imply a bit order.
It then propagates these bit orders to other word-level structures that have been reconstructed from the sea of gates but lack an ordering, such as registers.
Both the module identification plugin and the bit-order propagation plugin have been published alongside our ACM CCS 2024 publication~\cite{DBLP:conf/ccs/KlixASSVWLLKSHP24}.

After recovering word-level structures, the \textit{simulator} plugin enables the simulation of the netlist or parts thereof.
The analyst can assign values to the simulated gates and use the integrated waveform viewer to observe how the netlist behaves over time.
The simulation state at a user-selected point in time can also be visualized in the graph view by coloring the nets according to their logical value.
Verilator~\cite{verilator} is used for simulation, but other simulation engines can easily be integrated.
On top of simulation, the \textit{sequential symbolic execution} plugin can assist in reconstructing a netlist's high-level functionality by providing a semi-automated method of extracting word-level functional descriptions from a modularized netlist.

\subsection{Development}
The development of \HAL was initiated by Fyrbiak~\etal in 2018 with a first publication on the tool following in 2019~\cite{DBLP:journals/tdsc/FyrbiakWSHHWWTP19}.
Later that year, \HAL was released open-source by Wallat \etal~\cite{DBLP:conf/cf/WallatA00EFDMP19}.
Today, the \HAL project is actively maintained and developed at the \ac{MPI-SP} and at \ac{RUB}, with contributions from the community.
The core team includes PhD students, postdocs, and a dedicated full-time software developer.
Our \ac{CI} workflows rigorously test for functional correctness on supported platforms based on around \numprint{3500} manually designed test cases and automatically update our \ac{API} documentation.
Development funding comes exclusively from independent academic sources, initially ERC grant 695022 (EPoCH) and currently annual budgets at \ac{MPI-SP}.

\subsection{Documentation}
\HAL comes with extensive documentation that is continually expanded.
Automatically generated documentation covers the C++\footnote{\url{https://emsec.github.io/hal/doc/}} and Python \acp{API}\footnote{\url{https://emsec.github.io/hal/pydoc/}}.
Additionally, a manually maintained Wiki\footnote{\url{https://github.com/emsec/hal/wiki}} offers user-oriented guides and example projects demonstrating \HAL features, particularly the \ac{GUI} and Python interface.

%% file: sections/03_impact.tex
\section{Impact}
Although \HAL addresses a niche topic within a sensitive and sometimes legally complex area, it has made significant contributions to academic research and is beginning to gain adoption in industry and government contexts.

To date, we have identified 22 peer-reviewed publications and one preprint that utilize \HAL to varying degrees (see \autoref{tab:publications}).
These works can be categorized into 17 papers that utilize or advance \HAL for netlist analysis, 5 papers that leverage \HAL as a tool to study human behavior during hardware reverse engineering, and one paper that deploys \HAL for circuit optimization in the context of \acl{FHE} bootstrapping.
Without \HAL, several notable technical publications from our group---such as DANA~\cite{DBLP:journals/tches/AlbartusHTAP20}, HAWKEYE~\cite{DBLP:conf/crypto/LeanderPSS24}, research on breaking \ac{FSM} obfuscation~\cite{DBLP:journals/tches/FyrbiakWDABTP18}, and the reverse engineering of the iPhone~7 \ac{FPGA}~\cite{DBLP:conf/ccs/KlixASSVWLLKSHP24}---would not have been feasible within reasonable timeframe.
Many of these publications have also resulted in new open-source \HAL plugins as discussed in \autoref{hal::subsec::plugins}.

\begin{table}[htb]
    \centering
    \caption{Overview of academic publications involving \HAL.}
    \label{tab:publications}
    \begin{tabularx}{\linewidth}{L{5mm}|L{15mm}|X|L{6mm}}
        \toprule
        \textbf{Yr} & \textbf{Venue} & \textbf{Description} & \textbf{Ref} \\
        \midrule
        '24 & CCS & black-box \acs{FPGA} \acs{RE} & \cite{DBLP:conf/ccs/KlixASSVWLLKSHP24} \\
        '24 & CRYPTO & symmetric crypto analysis & \cite{DBLP:conf/crypto/LeanderPSS24} \\
        '24 & CHES & optimized bootstrapping in \acs{FHE} & \cite{DBLP:journals/tches/MonoKG24} \\
        '23 & HOST & targeted bitstream fault injection & \cite{DBLP:conf/host/EngelsEP23}\\
        '23 & ISQED & word-level \acs{RE} for \acsp{FPGA} & \cite{DBLP:conf/isqed/NarayananVPMV23} \\
        '23 & VLSID & word-level \acs{RE} for \acsp{FPGA} & \cite{DBLP:conf/vlsid/VenkatesanNPMV23} \\
        '23 & ISVLSI & \acs{FSM} \acs{RE} for \acsp{FPGA} & \cite{DBLP:conf/isvlsi/MuthukumaranVPNVE23} \\
        '23 & MWSCAS & \acs{FPGA} data-path \acs{RE} using \acsp{GNN} & \cite{DBLP:conf/mwscas/PulaVNMVE23}\\
        '23 & COINS & property-based Trojan detection & \cite{DBLP:conf/coins/ZakenMARM23} \\
        '23 & ePrint & proposing open-scan model & \cite{DBLP:journals/iacr/AzrielM23} \\
        '23 & TOCHI & problem solving strategies in \ac{HRE} & \cite{DBLP:journals/tochi/WiesenBWPR23} \\
        '22 & JCEN & attacks on logic locking & \cite{DBLP:journals/jce/EngelsHP22} \\
        '21 & CHES & \acs{HW} camouflaging for obfuscation & \cite{DBLP:journals/tches/HoffmannP21} \\
        '21 & CHES & framework for \acs{FPGA} protection & \cite{DBLP:journals/tches/StolzASKNGFPGT21}\\
        '20 & CHES & reconstruct registers & \cite{DBLP:journals/tches/AlbartusHTAP20} \\
        '20 & SOUPS & cognitive factors in \ac{HRE} & \cite{DBLP:conf/soups/0001WARP20} \\
        '19 & FIE & university-level course for \ac{HRE} & \cite{DBLP:conf/fie/Wiesen0APR19} \\
        '19 & ASP-DAC & cognitive obfuscation & \cite{DBLP:conf/aspdac/WiesenA0BWFRP19} \\
        '19 & ASP-DAC & \acs{FPGA} bitstream Trojan insertion & \cite{DBLP:conf/aspdac/EnderSWWKP19} \\
        '19 & CF & open-source release of \HAL & \cite{DBLP:conf/cf/WallatA00EFDMP19} \\
        '19 & TDSC & initial publication on \HAL & \cite{DBLP:journals/tdsc/FyrbiakWSHHWWTP19} \\
        '18 & CHES & breaking \acs{FSM} obfuscation & \cite{DBLP:journals/tches/FyrbiakWDABTP18}\\
        '18 & TALE & skill acquisition in \ac{HRE} & \cite{DBLP:conf/tale/WiesenBFAERP18}\\
         \bottomrule
    \end{tabularx}
\end{table}

Given the sensitivity of hardware reverse engineering and its applications, we cannot disclose companies or government entities using \HAL.
However, the framework's impact on these stakeholders is reflected in its accumulation of 680 GitHub stars and 86 forks over the years, alongisde the success of dedicated hardware and \ac{FPGA} reverse engineering trainings and workshops.
For instance, we taught trainings at \textit{hardwear.io} from 2023 to 2025 and at the \textit{Summer School on Real-World Crypto and Privacy} in 2022 and 2025, personally instructing more than 150 professionals in the practical use of \HAL.
Furthermore, since 2017, we have lectured well-attended university courses on hardware reverse engineering at \acl{RUB} and UMass Amherst.
These courses, built around continuously evolving versions of \HAL, have taught approximately 200 students.

In addition to formal education, we have publicly presented and demonstrated \HAL on multiple occasions, including an invited talk at \ac{FSIC} 2025~\cite{klix2025hal} and a selected talk at the \ac{36C3}~\cite{hoffmann2019hal}.
Currently, we are exploring pathways to transition \HAL into a commercial technology that extends beyond reverse engineering to support hardware design and verification.

%% file: sections/04_conclusion.tex
\section{Conclusion}
\HAL fills a critical gap in \acl{HRE} by providing an open-source, extensible framework that advances research, education, and practical analysis of gate-level netlists. 
Its wide adoption---demonstrated by numerous academic publications, integration into university courses, and industry workshops---attests to its value and impact on the community. 
By enhancing accessibility and reproducibility, \HAL empowers analysts to better understand and mitigate hardware vulnerabilities. 
Ongoing development and community engagement ensure that \HAL will continue to foster innovation and collaboration in this sensitive yet vital cybersecurity field.